\documentclass[twocolumn,showpacs,preprintnumbers,amsmath,amssymb,superscriptaddress,groupedaddress,prl]{revtex4}

\usepackage{epsfig}                                                            
\usepackage{graphicx}
\usepackage{dcolumn}
\usepackage{color}
\usepackage{bm}
\usepackage{subfigure} 
\usepackage{epsfig} 

\begin{document}

\title{\bf Superspin Glass Mediated Giant Spontaneous Exchange Bias in a Nanocomposite of BiFeO$_3$-Bi$_2$Fe$_4$O$_9$}

\author {Tuhin Maity} \affiliation {Micropower-Nanomagnetics Group, Microsystems Center, Tyndall National Institute, University College Cork, Lee Maltings, Dyke Parade, Cork, Ireland}
\author {Sudipta Goswami} \affiliation {Nanostructured Materials Division, CSIR-Central Glass and Ceramic Research Institute, Kolkata 700032, India}
\author {Dipten Bhattacharya} 
\email{dipten@cgcri.res.in} \affiliation {Nanostructured Materials Division, CSIR-Central Glass and Ceramic Research Institute, Kolkata 700032, India}
\author {Saibal Roy}
\email{saibal.roy@tyndall.ie} \affiliation {Micropower-Nanomagnetics Group, Microsystems Center, Tyndall National Institute, University College Cork, Lee Maltings, Dyke Parade, Cork, Ireland} 

\date{\today}

\begin{abstract}
We observe an enormous $\textit{spontaneous}$ exchange bias ($\sim$300-600 Oe) - measured in an unmagnetized state following zero-field cooling - in a nanocomposite of BiFeO$_3$ ($\sim$94\%)-Bi$_2$Fe$_4$O$_9$ ($\sim$6\%) over a temperature range 5-300 K. Depending on the path followed in tracing the hysteresis loop - positive (p) or negative (n) - as well as the maximum field applied, the exchange bias ($H_E$) varies significantly with $\mid-H_{Ep}\mid$ $>$ $\mid H_{En}\mid$. The temperature dependence of $H_E$ is nonmonotonic. It increases, initially, till $\sim$150 K and then decreases as the blocking temperature $T_B$ is approched. All these rich features appear to be originating from the spontaneous symmetry breaking and consequent onset of unidirectional anisotropy driven by "superinteraction bias coupling" between ferromagnetic core of Bi$_2$Fe$_4$O$_9$ (of average size $\sim$19 nm) and canted antiferromagnetic structure of BiFeO$_3$ (of average size $\sim$112 nm) via superspin glass moments at the shell. 
\end{abstract} 
\maketitle
 
The spontaneous exchange bias (SEB), where the unidirectional anisotropy (UA) sets in $\textit{spontaneously}$ under the application of the first field of a hysteresis loop even in an unmagnetized state, is a consequence, primarily, of biaxial symmetry in the antiferromagnetic (AFM) structure of ferromagnetic (FM)/AFM interface \cite{Saha,Wang,Ahmadvand}. In a spin glass (SG)/FM structure, on the other hand, the anisotropy sets in under field cooling via oscillatory RKKY interaction \cite{Ali}. However, we show in this Letter that glassy moments at the interface, in fact, introduce an additional magnetic degree of freedom in between the exchange-coupled FM and AFM grains and breaks the symmetry truly spontaneously even before the application of the first field of a loop to set the UA in an unmagnetized state. As discussed later, the consequence of this is an asymmetry in the SEB depending on the path followed in tracing the hysteresis loop - positive or negative. We report that in a nanocomposite of BiFeO$_3$($\sim$94\%)-Bi$_2$Fe$_4$O$_9$ ($\sim$6\%), we observe (i) a large SEB ($\sim$300-600 Oe) across 5-300 K, (ii) asymmetry in SEB depending on the path followed in tracing the hysteresis loop - positive or negative, and (iii) a nonmonotonic variation of SEB with temperature. The magnitude of the SEB itself is far higher than what has so far been observed in all the bulk or thin film based composites of BiFeO$_3$ \cite{Martin,Ramesh,Chu,Heron,Lebeugle,Sahoo} even under magnetic annealing. We have also observed the conventional magnetic-annealing-dependent exchange bias (CEB) with all its regular features such as dependence on annealing field, rate, and training. The random field generated by the glassy moments at the shell appears to be influencing the indirect exchange bias coupling called "superinteraction bias coupling" between FM core \cite{Tian} of finer ($\sim$19 nm) Bi$_2$Fe$_4$O$_9$ and local moments of AFM order in coarser BiFeO$_3$ ($\sim$112 nm) and inducing the SEB, its path dependence, and its nonmonotonicity in variation with temperature. 

\begin{figure}[!h]
  \begin{center}
    \includegraphics[scale=0.45]{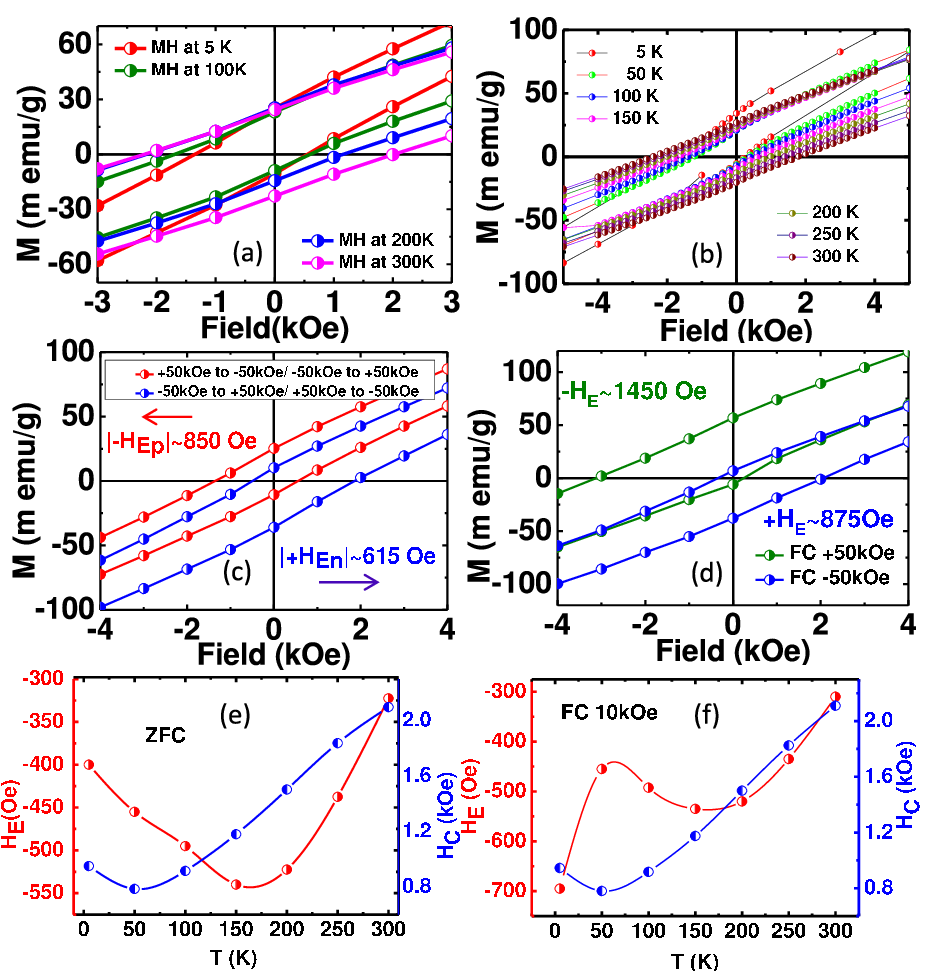} 
    \end{center}
  \caption{(color online) (a) The hysteresis loop shift, signifying SEB at different temperatures across 5-300 K, measured under 50 kOe following zero-field cooling; the region near origin of the loop is blown up to show the extent of exchange bias clearly. (b) The CEB at different temperatures across 5-300 K measured under a field cooling with +10 kOe. (c) The switch in sign and change in magnitude of the loop shift at 5 K signaling asymmetry and tunability of the SEB depending on the sign of the starting field (+50 kOe/-50 kOe) of hysteresis loop measurement. (d) The switch in sign and change in magnitude of the CEB at 5 K measured following field cooling under +50 kOe/-50 kOe. (e) The variation of SEB and corresponding H$_C$ with temperature. (f) The variation of CEB - measured following field cooling under +10 kOe - and corresponding H$_C$ with temperature (lines are guide to the eyes).}
\end{figure}

The nanocomposite of BiFeO$_3$-Bi$_2$Fe$_4$O$_9$ has been synthesized by the sonochemical route \cite{Goswami}. By varying the processing conditions such as temperature, time, atmosphere, etc., the volume fraction of the Bi$_2$Fe$_4$O$_9$ phase has been varied from $\leq$3\% to $\sim$10\%. The volume fraction of the phases, crystallographic details of each phase, particle morphology, average misalignment angle between two single crystalline nanoparticles of the component phases, etc., have been determined from rigorous x-ray diffraction, transmission electron microscopy, selected area and convergent beam electron diffraction experiments \cite{supplementary}. The magnetic measurements have been carried out in a SQUID magnetometer (MPMS, Quantum Design) across 5-300 K under a maximum field H$_m$ of 50 kOe. In order to ensure that there is no trapped flux both in the superconducting coil of MPMS and in the sample we followed a well-designed protocol to demagnetize them. The superconducting coils of MPMS are normally discharged from high field (50 kOe) in oscillation mode; the amount of trapped flux is typically $\sim$10 Oe. Before starting a new batch of experiment, the magnet was warmed to room temperature which is above the critical point. In addition, prior to each measurement, the sample itself was demagnetized with oscillating field using an appropriate protocol (given in the Supplemental Material\cite{supplementary}) in order to ensure that there is no trapped flux in the sample. The details of the protocol as well as the results of measurement on diamagnetic sample (sapphire) have been given in the Supplemental Material \cite{supplementary}. We have also measured the SEB at 300 K for a maximum field of 18 kOe following zero-field cooling (ZFC) from a high temperature ($\sim$700 K) - which is even above the magnetic transition point T$_N$ ($\sim$590 K) of the AFM component - in a vibrating sample magnetometer for a test case \cite{supplementary}. We obtain an SEB of $\sim$81 Oe at 300 K which is consistent with the SEB \cite{Maity} for different H$_m$ across 10-50 kOe measured in SQUID. This shows that the demagnetization protocol used in SQUID was appropriate in ensuring unmagnetized state of the sample prior to the measurement.    

\begin{figure}[!h]
  \begin{center}
    \includegraphics[scale=0.55]{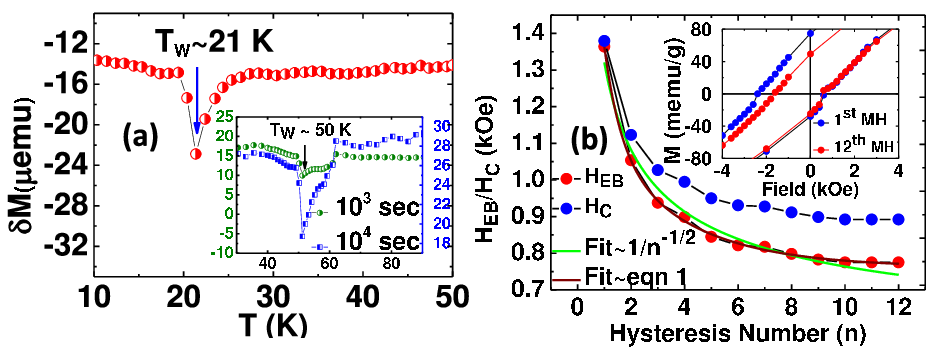} 
    \end{center}
  \caption{(color online)  (a) The characteristic dip at $\sim$21 K in the differential between two ZFC magnetization versus temperature patterns recorded under two different protocols - a simple ZFC and a ZFC with "stop-and-wait" approach; inset shows a similar dip even at 50 K. It appears to become sharper and more prominent with the increase in wait time. (b) The impact of training effect on CEB for sample A. The CEB and $H_C$ decrease with the increase in number of hysteresis cycles (n); inset shows a portion of the loop at first and twelfth cycle.}
\end{figure}

\begin{figure*}[!htp]
  \begin{center}
	 \includegraphics[scale=0.65]{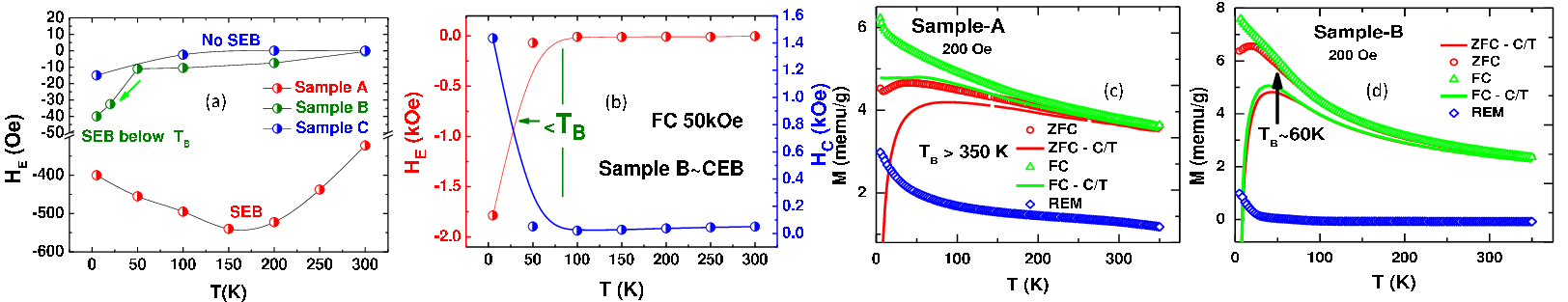} 
    \end{center}
  \caption{(color online) (a) The SEB for all the three samples with different volume fractions of Bi$_2$Fe$_4$O$_9$ phase. (b) The CEB and H$_C$ versus temperature plot for sample-B. Large CEB (measured following field cooling under 50 kOe) could be observed at only below $T_B$. The zero-field cooled (ZFC), field cooled (FC), and remanent magnetization versus temperature plots for (c) sample A and (d) sample B. The solid lines show the ZFC and FC magnetizations after subtraction of the contribution of the paramagnetic C/T component in both the cases; $T_B$ turns out to be $>$350 K for sample A and $\sim$60 K for sample B.}
\end{figure*}

\begin{figure}[!h]
  \begin{center}
    \includegraphics[scale=0.40]{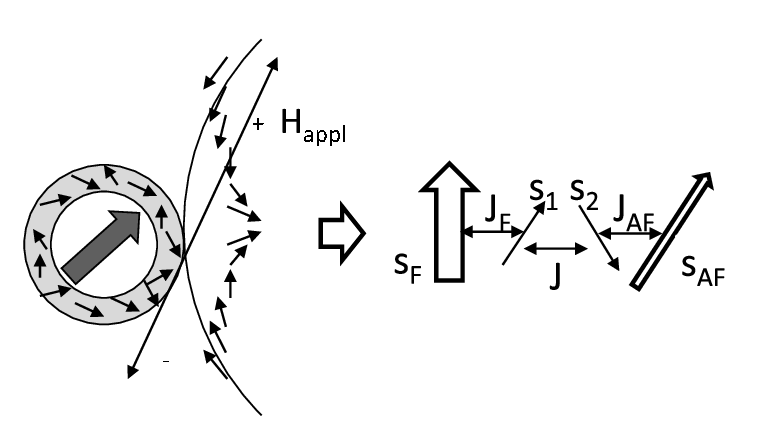} 
    \end{center}
  \caption{Schematic of the ferromagnetic and antiferromagnetic spin interaction via superspin glass moments at the interface; left part shows the ferromagnetic core of finer Bi$_2$Fe$_4$O$_9$ particle and superspin glass moments at the shell interacting with the local moments of spiral spin spin structure of bigger BiFeO$_3$; right part shows the spin configuration and interaction energies. }
\end{figure}

We report here mainly the results obtained in a nanocomposite of $\sim$6\% Bi$_2$Fe$_4$O$_9$ and $\sim$94\% BiFeO$_3$ (sample A) which exhibits maximum SEB and CEB. In Fig. 1, the results from the magnetic measurements are shown. In Fig. 1(a), we show the hysteresis loops which yield the SEB at several temperatures across 5-300 K. The region near the origin is blown up to show the extent of EB clearly (full loops are given in the Supplemental Material \cite{supplementary}). We used a field step size of 100 Oe near the origin of the hysteresis loop in order to measure the exchange bias accurately. The field span of 10 kOe under such a protocol is covered typically within $\sim$3 h ($\sim$10$^4$s) which gives the time scale of the measurement. In each case, the presence of a large shift in the loop along the field axis is conspicuous. This shift cannot result from relaxation of coercivity of the FM component as the tensorial nature of the magnetocrystalline anisotropy cannot contribute to the unidirectional anisotropy. The EB $H_E$ is given by ($H_{c1}$ + $H_{c2}$)/2 while the coercivity $H_C$ is given by ($H_{c1}$ - $H_{c2}$)/2; $H_{c1}$ and $H_{c2}$ are the fields (signs are included) corresponding to the points in forward and reverse branches of the hysteresis loop at which the magnetization reaches zero. The extent of SEB observed here right across 5-300 K is quite large and comparable to what has been reported by Wang $\textit{et al}$. \cite{Wang} in Ni-Mn-In bulk alloys at 10 K. While ramping the temperature from one point to another, a constant ramp rate of 2.5 K/min has been used. The observation of SEB iteslf in BiFeO$_3$ based bilayer or composite system has not been reported so far, and, for the first time, we are reporting it in the nanocomposite of BiFeO$_3$-Bi$_2$Fe$_4$O$_9$. In Fig. 1(c), the asymmetry and hence the $\textit{tunability}$ of the SEB at 5 K has been demonstrated. Depending on the sign of the starting field +50 kOe(-50 kOe), the sign of the SEB is negative (positive) as well as $\mid$-H$_{Ep}$$\mid$ $>$ $\mid$+H$_{En}$$\mid$. This is also remarkable and has not yet been observed in any other system exhibiting SEB \cite{Wang}. Figure 1(b) shows the CEB measured after a magnetic annealing treatment with 10 kOe. In this case a field of 10 kOe has been applied at room temperature and then the temperature was ramped down to the given point at a cooling rate of 2.5 K/min. Like SEB, the CEB too turns out to be negative i.e., annealing under positive (negative) field yields hysteresis loop shift in negative (positive) direction along the field axis. Even more interesting is that, in this case too, the exchange bias $H_E$ for positive (negative) annealing field is $\textit{asymmetric}$ with $\mid$-H$_{Ep}$$\mid$ $>$ $\mid$+H$_{En}$$\mid$. This has been demonstrated clearly in Fig. 1(d) which shows the asymmetry in the shift of the loop along the field axis at 5 K depending on whether the sample has been field-cooled under +50 kOe or -50 kOe. In Figs. 1(e) and 1(f), respectively, we show the $H_E$ and $H_C$ as a function of temperature (T) for SEB (ZFC) and CEB (measured following FC with 10 kOe). The H$_E$ and H$_C$ in both of these cases are nearly identical in magnitude and nonmonotonic. While H$_E$-T plots exhibit valleys at $\sim$150 K for both SEB and CEB, the H$_C$-T plots exhibit valleys at $\sim$50 K. In addition, the H$_E$-T plot exhibits a peak at $\sim$50 K for CEB [Fig. 1(f)]. The nearly identical magnitude of H$_E$ and H$_C$ signifies nearly identical UA and domain pinning under ZFC and FC with 10 kOe. $H_E$, however, is large at 5 K, possibly, because of large magnetization at low temperature which could increase further under field cooling.       

In order to trace the origin of all these features, we investigated the spin structure both in the bulk of the BiFeO$_3$ and Bi$_2$Fe$_4$O$_9$ particles as well as at their interfaces from well designed protocol dependent magnetic memory and training effect measurements. We obtained a profound signature of the presence of superspin glass (SSG) moments in the memory effect measurement for sample A. We used a 'stop-and-wait' protocol to measure the memory effect which is an unequivocal signature of the presence of SSG \cite{Sasaki}. The sample was first cooled down to 2 K from room temperature under zero field and an M(T) pattern (which acts as reference line) was measured under 200 Oe. After the sample temperature reaches 300 K, it was again brought back to 2 K under zero field. The M(T) measurement was then repeated but with a 'stop-and-wait' protocol. As the temperature reaches at $T_w$ $\sim$21 K, the measurement was stopped for $\sim$10$^4$s and the restarted to reach back to 300 K. The difference between the two patterns $\delta$M(T) is shown in Fig. 2(a), the main frame. The memory effect is shown as a dip at $\sim$21 K which confirms the presence of SSG phase in the nanocomposite. The entire measurement has been repeated for $T_w$ $\sim$50 K [Fig. 2(a) inset]. The memory effect could be observed even at $\sim$50 K as well. We further measured the wait-time dependence of the memory effect [Fig. 2(a) inset]. It appears that the effect becomes sharper and more prominent with the increase in wait time across 10$^3$-10$^4$s. The SSG moments develop due to interaction among the frozen superparamagnetic domains - possibly present at the shell of the finer Bi$_2$Fe$_4$O$_9$ particles of core-shell structure with FM core - at finite interparticle distance below the blocking temperature ($T_B$ $>$ 350 K for sample A) \cite{Chen}. With the rise in exchange coupling strength, the superparamagnetic particles form a SSG, initially, and then even a superferromagnetic phase.  

The dynamics of the spin structure at the interface has been probed for sample A by studying the training effect on CEB at 5 K for 12 repeating cycles. The dependence of $H_E$ and $H_C$ on the number of repeating cycles (n) is shown in Fig. 2(b). The CEB obtained under a H$_m$ of 50 kOe following FC with 50 kOe is shown here. Both the parameters are found to be decreasing monotonically with the increase in n indicating spin rearrangement at the interface. It appears that the empirical law \cite{Paccard} for purely AFM spin rearrangement at the interface $H_E^n$ = $H_E^\infty$ + k.n$^{-\frac{1}{2}}$ with k = 505 Oe and $H_E^\infty$ = 813 Oe cannot describe our data well [green line in Fig. 2(b)]. Instead, a model \cite{Mishra} which considers a mixed scenario of two different relaxation rates for frozen and rotate-able uncompensated spin components at the interface 
\begin{equation}
H_E^n = H_E^\infty + A_fexp(-n/P_f) + A_rexp(-n/P_r)
\end{equation}
(where f and r denote the frozen and rotate-able spin components) fits the data perfectly well [brown line in Fig. 2(b)] and yields the fitting parameters as $H_E^\infty$ = 761 Oe, $A_f$ = 1394 Oe, $P_f$ = 0.61, $A_r$ = 451 Oe, and $P_r$ = 3. The ratio $P_r$/$P_f$ $\sim$ 5 indicates that the rotate-able spins rearrange nearly 5 times faster than the frozen spins. Thus while the 'memory effect' signifies the presence SSG moments in the nanocomposite, the 'training effect' on CEB shows that the SSG moments reside at the interfaces between FM Bi$_2$Fe$_4$O$_9$ and AFM BiFeO$_3$ particles and influence the SEB and CEB significantly. It is important to mention here that the SEB exhibits negligible training effect within the laboratory time scale ($\sim$10$^4$s). This could be because it originates from a stable state under zero field and zero magnetization through spontaneous symmetry breaking.
 
We further examined the SEB in two other samples with higher ($\sim$10\%) and lower ($<$3\%) volume fraction of Bi$_2$Fe$_4$O$_9$ (sample B and C, respectively). The corresponding full hysteresis loops have been given in the Supplemental Material \cite{supplementary}. The T$_N$ of the AFM component for sample B and C are $\sim$490 K and $\sim$450 K, respectively (given in the Supplemental Material \cite{supplementary}). In Fig. 3(a), comparison of the SEB among all the three samples (A,B, and C) is shown. The SEB is found to follow a rather nonmonotonic pattern with the variation in the volume fraction of Bi$_2$Fe$_4$O$_9$ phase. It decreases both with the increase and decrease in the volume fraction of the Bi$_2$Fe$_4$O$_9$ phase. The SEB in all these cases could be observed at only below the respective $T_B$'s. The $T_B$ decreases down to $\sim$60 K in sample-B because of finer Bi$_2$Fe$_4$O$_9$ particles ($\sim$8 nm). The $T_B$, however, could not be located within the range 5-300 K for sample C and, therefore, no exchange bias could be observed in this sample within the same temperature range. The CEB and $H_C$ for sample B are also found to be finite [Fig. 3(b)] only at below the $T_B$ ($\sim$60 K). And as expected, the memory effect too has been observed in sample B at below $T_B$ \cite{supplementary}.  The memory effect, observed both in sample A and B, implies presence of SSG phase and its influence on the exchange bias. Since superparamagnetic and SSG phases coexist at below $T_B$ in both the samples, one can estimate the relative volume fraction of the SSG phase by calculating the ZFC and FC magnetic moment versus temperature pattern after subtracting the contribution of the Curie paramagnetic component C/T (C = Curie constant) from the experimental result [Figs. 3(c) and 3(d)] and noting the flatness of the FC moment versus temperature pattern \cite{Chen} at below $T_B$. The calculated patterns (solid lines) for both the samples A and B are shown in Figs. 3(c) and 3(d), respectively. It appears that the volume fraction of SSG phase is higher in sample A than in sample B. The SEB too is higher in sample A than in sample B. Clear correlation between the volume fraction of the SSG phase and the extent of SEB shows that the SSG phase plays a major role in inducing SEB. 

We show that all these results could be qualitatively understood by considering a model of "superinteraction bias coupling" between the FM core of finer Bi$_2$Fe$_4$O$_9$ and local uncompensated moments of the AFM order in coarser BiFeO$_3$ particles via the SSG shell at the interface. The model is shown schematically in Fig. 4 and draws essentially from the model proposed in Ref. [3]. The dotted line marks the direction of the applied field. The shell SSG moments $s_1$ and $s_2$ are coupled to the FM moment $S_F$ by a coupling parameter $J_F$ and to the AFM moment $S_{AF}$ by $J_{AF}$ while the coupling between $s_1$ and $s_2$ is $J$. The net coupling parameter $b$ will depend on $J_{AF}$, $J_F$, and $J$ and, finally, $H_E$ $\propto$ $b$ \cite{Ali}. It has been shown \cite{Ali} that the random fields generated by spin glass moments at the core can act on the saturated FM moment and set the UA via RKKY interaction. The model that we are proposing in the present case is the following. The random field from frozen SSG moments appears to be inducing a variation in the anisotropy of the AFM moments including biaxiality with respect to the direction of the applied field. Thus depending on the orientation of the principal easy axes of AFM grains with respect to the direction of the applied field, the AFM grains can experience either no torque or large torque and become (i) fully hysteretic, (ii) non-hysteretic, or (iii) partially hysteretic. While the fully hysteretic and non-hysteretic grains do not contribute to the bias in the loop, the partially hysteretic grains do. The partially hysteretic grains set the UA, primarily, in a direction opposite to that of the applied field. The SEB, then, becomes negative - i.e., depending on the sign of the starting field for loop tracing, positive (negative), the SEB turns out to be negative (positive). Application of the first field for tracing the loop breaks the symmetry among the AFM grains and sets the UA. The FM moments are assumed to be saturated under the applied field. However, the most interesting aspect is that there is a $\textit{spontaneous symmetry breaking}$ as well, driven by the random field of the SSG moments at the interface which yields a global minima in the energy landscape and sets the UA universally along the negative field direction even in absence of first field of loop tracing. These grains are thus always partially hysteretic along the negative direction of the applied field. The grains which set the UA in a direction opposite to that of applied field are partially hysteretic for both the directions of applied field. But the ones mentioned above are partially hysteretic $\textit{only}$ with respect to the negative field direction. This aspect, in fact, gives rise to the observed $\textit{asymmetry}$ in both SEB and CEB with $\mid$-H$_{Ep}$$\mid$ $>$ $\mid$+H$_{En}$$\mid$ and has not been reported by others so far in the context of either SEB or CEB. The role of SSG moments, therefore, appears to be crucial in inducing this spontaneous symmetry breaking and setting the UA universally along the negative field direction. Alternatively, a similar effect could be observed due to an even finer fraction of Bi$_2$Fe$_4$O$_9$ particles, because of a distribution in the size, which form superferromagnetic (SFM) domains via stronger interparticle exchange interaction \cite{Chen}. The SSG mediated SFM-AFM exchange interaction within an ensemble of grains with a finer fraction of Bi$_2$Fe$_4$O$_9$ particles, in that case, could actually give rise to the spontaneous symmetry breaking and set the UA universally along the negative field direction even in absence of first field of loop tracing. Only those grains, then, are responsible for giving rise to the observed asymmetry in SEB and CEB. 

The temperature dependence of SEB is nonmonotonic as at well below $T_B$, the increase in temperature increases the interaction between SSG and AFM moments which, in turn, induces the energy landscape necessary to set the UA in the system. The bias as well as the asymmetry, therefore, increase. However, as the $T_B$ is approached, the number of grains turning superparamagnetic increases which, in turn, reduces the bias. The nonmonotonic variation in SEB with the volume fraction of Bi$_2$Fe$_4$O$_9$ phase, likewise, can be explained by considering nonmonotonic variation in the volume fraction of the SSG phase.           

In summary, we report a giant as well as tunable spontaneous exchange bias of $\sim$300-600 Oe across 5-300 K in a nanocomposite of BiFeO$_3$ ($\sim$94\%) - Bi$_2$Fe$_4$O$_9$ ($\sim$6\%). It originates from a superinteraction bias coupling between ferromagnetic core of finer Bi$_2$Fe$_4$O$_9$ ($\sim$19 nm) particles and antiferromagnetic moment in coarser ($\sim$112 nm) BiFeO$_3$ particles via superspin glass moments at the interface. Since it induces a variety of coupling across the interfaces and thus develop a complicated interaction energy landscape among the FM/AFM grains by breaking the symmetry spontaneously even before the application of the first field of the loop tracing, the presence of superspin glass moments turns out to be crucial. This giant and tunable (i.e., path dependent) exchange bias can be utilized for an enormous improvement in the efficiency of switching the magnetic anisotropy in a ferromagnetic system electrically via "exchange coupling mediated multiferroicity" in such a nanocomposite and/or a multilayer thin film systems. 

This work has been supported by Indo-Ireland joint program (DST/INT/IRE/P-15/11), Science Foundation of Ireland (SFI) Principal Investigator (PI) Project No. 11/PI/1201 and FORME SRC project (07/SRC/I1172) of SFI. One of the authors (S.G.) acknowledges support from a Research Associateship of CSIR.

\end{document}